# Impact of ionization of ferrocene: EOES of α- and β-electrons and the fingerprint orbital 8a$_1$' of ferrocenium

Feng Wang[*] and Shawkat Islam

Molecular Model Discovery Laboratory, Department of Chemistry and Biotechnology, Faculty of Science, Engineering and Technology, Swinburne University of Technology, Hawthorn, Melbourne, Victoria, 3122, Australia

*Corresponding author. Tel.: +61 3 9214 5065; Fax: +61-3-9214-5921.

E-mail addresses: fwang@swin.edu.au (F. Wang).

(15/09/2014)

**Abstract**

Ionization process of ferrocene (Fc) to produce ferrocenium cation (Fc$^+$) has been debated as much as the eclipsed and staggered ferrocene conformers. The present quantum mechanical study reveals that removal of an electron does not apparently affect the geometry and symmetry of the cation, as the geometric changes are < 2% with respect to neutral Fc, but produce the fingerprint orbital 8a$_1$' of Fc$^+$. The excess orbital energy spectrum (EOES) of the α- and β-electrons reveals that the electrons originated from the transition metal Fe in both core and valence shells experience significant energy changes in the cation with respect to the neutral ferrocene counterparts, indicating that the Fe-electrons correlate stronger than electrons from other atoms such as carbons in ferrocene. The EOES also exhibits that the orbital energies of the α-electrons in ferrocenium change more significantly than the β-electrons after one β-electron being ionized with respect to ferrocene. The most significant changes upon ionization are dominated by the Fe-electrons in non-degenerate signature orbitals. That is, $4a_1'$ which is dominated by the Fe 3s orbital, $3a_2''$ by the Fe $3p_z$ orbital and the $8a_1'$ orbital (α only) by the Fe $3d_z^2$ (one β-electron is removed from this orbital). The ΔDFT and ΔSCF calculations yield the IP of 6.90 eV and 6.85 eV, respectively, in excellent agreement to the recent measurements of 6.9±0.1 eV, suggesting that significant relaxation energy exists whereas electron correlation energies are largely cancelled out. A further dual space analysis (DSA) identifies that the momentum profiles of such the singly occupied fingerprint $8a_1'$ orbital indeed experiences most significant changes in all orbitals of $Fc^+$, again indicates that the electron removal of ferrocene ionization is not from the highest occupied molecular orbital.

## 1. Introduction

The significance of the discovery of di-cyclopentadienyl iron (η5, $FeCp_2$) i.e., ferrocene (Fc),[1] is not only a landmark for organometallic chemistry with an immense number of applications, but also in its own right for the study of chemical bonding and electron correlation models. The uniquely high thermal stability of ferrocene breaks the general view that the transition metal-carbon bonds were very unstable. Various coordination modes of hydrocarbon ligands are developed since the industrial importance of organometallic compounds of transition metals increased with the discoveries of a variety of catalysts has been recognised.

Ionization process with one β-electron being removed from the neutral ferrocene (Fc) to produce ferrecenium cation (Fc$^+$) has been debated as much as the eclipsed (D$_{5h}$) and staggered (D$_{5d}$) ferrocene conformers, both theoretically and experimentally.[2] A number of questions regarding the ionization of ferrocene remain to be answered. For example, what is the ground electronic state of ferrocenium, is it X$^2$A$_1$' or X$^2$E$_2$' (for the eclipsed conformer)? Would an electron from the highest occupied molecular orbital (HOMO) always be removed first at ionization? If not, why? How large is the relaxation energy at ferrocene ionization? How differently the electrons of the transition metal Fe correlate with other electrons in ferrocene and ferrocenium? Would the ionization of ferrocene only impact on outer valence electrons? How different the α-electrons and β-electrons respond to the removal of one β–electron of ferrocene?

Although significant advances in the development of experimental techniques have been employed to study Fc, the information obtained so far is still insufficient to understand its unique structures and properties of Fc without theoretical insight. As indicated by Coriani et al[3], Fc is a "notoriously difficult example" as it contains transition metal Fe which leads to much larger errors because of more complex bonding situations and its d-electrons. In addition, the fact that ferrocene is a Jahn-Teller effect inactive molecule,[4] also contributes to this challenge. As a result, previous studies of Fc and Fc$^+$ are largely contradictory.[5-7] Many recent quantum mechanical studies using high level post-HF methods, such as MP2, CCSD and CCSD(T) in combination with large basis sets such as TZV2P+f[3] and various DFT models including BHLP,

B3LYP, BLYP, BP86, LSDA[8] and BPW91[9] models suggested that the eclipsed Fc is likely the global minimum structure of Fc in its ground electronic state ($S_1$).[3,10-13]

The recent DFT study[10] accurately predicted the signature vibrational transitions at 400-500 cm$^{-1}$ of the infrared (IR) region, which brings a clue to identify ferrocene conformers in addition to their relative energies.[10] Moreover, a more recent excess orbital energy spectrum (EOES) of ferrocene conformers[14] reveals that the electrons of transition metal Fe play different roles in the conformers of ferrocene. Larger exceed orbital energies have been discovered in the Fe related core and inner valence shell of the complex from the EOES.[14] The study indicates that a group of Fe-dominant/associate orbitals of Fc exhibit larger excess orbital energies, some of which are even larger than the total electron energy difference between the E-Fc and S-Fc conformers.[14]

It is well recognised in quantum chemistry that in addition to the level of theory employed, the basis sets play a very important role in the Pople diagram.[15] The properties of ferrocene particularly demonstrate the need for the appropriate basis set of the transition metal Fe. For example, the CASPT2 (complete active second-order perturbation theory) and CCSD(T) (coupled cluster) methods are now considered as among the most accurate methods in quantum chemistry,[16] the calculated binding energy of ferrocene results to a value which is 20 kcal·mol$^{-1}$ different from experiment.[17-18] This result[16-17] in fact takes considerations of all possible corrections such as electron correlation energies, relativistic effects, basis set incompleteness, 3s/3p semi-core correlation, zero-point energy and thermal correction to enthalpy and geometry relaxation of both the Cp$^-$ rings and the Fc molecule. Unfortunately, the resulting binding energy of ferrocene is considered as unsatisfactory as the undisputed adequacy of these methods to achieve chemical accuracy.[16-17] An alternative study discovered that, for transition metals such as Fe, both atomic configurations of $3d^n4s^1$ and $3d^{n-1}4s^2$ are important and must be included in the basis set,[19] in order to more appropriately describe the Fe-contained complexes such as ferrocene.[10,14]

In the present study, we provide solid quantum mechanical analysis to understand the behaviour of α- and β-electrons of ferrocenium with respect to ferrocene. Only the eclipsed ($D_{5h}$) conformer of Fc is discussed in the present study, since several prior experimental[6,20-22] and theoretical studies[2,3,8-10,23-25] have established that the eclipsed conformer is more stable or the equilibrium conformer of Fc (and $Fc^+$). It is discovered that the $D_{5d}$ conformer of Fc behaves similarly in this regard. The orbitals with significant changes in $Fc^+$ are further studied using dual space analysis (DSA)[26] in coordinate space (orbital density distributions) and in momentum space (orbital momentum profiles).

## 2. Computational methods

In this study, unrestricted Hartree-Fock (HF) and density functional theory (DFT) based unrestricted B3LYP methods are employed for the calculations of the $Fc^+$ cation, along with the recently developed basis set for the transition metal Fe, that is, the m6-31G(d) basis set.[19] Due to the incorporation of necessary diffuse d-type functions for the central transition metal Fe, this basic set is able to describe the important energy differences between the atomic $3d^n4s^1$ and $3d^{n-1}4s^2$ configurations more precisely than the conventional 6-31G(d) basis set and a number of other basis sets for the central metal atom of Fc.[19,24] The 6-31G(d) basis set is employed for C and H atoms in Fc without modification. This modified basis set for Fe is able to produce the most accurate infrared (IR) spectra *without any scaling* for ferrocene in conjunction with the B3LYP functional.[10] As pointed by Phung et al[27], among a number of functionals including PBE0 and M06, the B3LYP functional performs best in reproducing the binding energy, so that the B3LYP method has been served as a set benchmark reference along with other high-level ab initio approaches.[28] Furthermore, our recent study proved that the B3LYP/m6-31G(d) model is able to distinguish between Fc conformers through the excess orbital energy spectrum (EOES).[14] In addition, the B3LYP functional was found to calculate the accurate shape of molecular orbitals.[26] All calculations using other quantum mechanical models in the present study are based on the optimized geometry of eclipsed Fc/Fc$^+$ using the B3LYP/m6-31G(d) model (Fc) and the UB3LYP/m6-31G(d) model (Fc$^+$).

The present study calculates the α-EOES and β-EOES between ferrocenium and ferrocene on an orbital base, where the symmetry of ionized system is correlated based on that for the neutral Fc system. The impact of other possible factors including relativistic effects and long range dispersion forces[30-33] etc may be insignificant due to the cancellation using the EOES. As indicated by Salzner[13] that there was no obvious change in the results with relativistic pseudo-potentials.[34-35] The one-on-one correspondence of the irreducible representations of α- and β-electrons of both ferrocenium (Fc$^+$) and ferrocene (except for the ionized β-electron) makes it possible to perform their EOESs. The α-EOES and β-EOES for Fc$^+$/Fc show how different the α-electrons and/or the β-electrons of Fc$^+$ from Fc. The electrons in the neutral Fc serve as the reference to monitor the changes of the electrons upon oxidation (ionization), based on the (U)B3LYP/m6-31G(d) and the (U)HF/m6-31G(d) models. Some selected orbitals of Fc$^+$ with significant changes from their Fc counterparts are further studied on their electron density distribution (theoretical) and orbital momentum profiles using dual space analysis.[26] The Gaussian09 computational chemistry package[36] has been employed to calculate all the results used in the present study. Theoretical momentum distribution calculations were performed using NEMS package.[37]

3. Results and discussion

*Minor geometric changes of bonds involve Fe in ferrocenium (Fc$^+$)*

Several previous studies have already indicated that for DFT calculations, the models yielding the lowest total electronic energy may not be the most accurate model for high level calculations of the molecular properties such as orbital shape.[26,29,38] This is particularly true for non-variational methods and organometallic compounds involving transition metals. It is well established that the geometric parameters as well as several other properties of Fc are found to be dependent on both the level of theory and basic set.[13,39] The earlier IR spectral calculation[10] and the recent EOES study[14] reveal the pivotal contribution of the central transition metal Fe in the Fc complex. Therefore, in addition to the level of theory, the basic set appropriately describing the transition metal Fe becomes very important to achieve the accuracy in the Pople diagram. As pointed out by Chiodo et al[33] that a remarkable tendency to stabilize atomic and cationic configurations with more d-type orbital occupation for transition metals such as Fe. The d-type diffuse basis function seems to play a very important role for the determination of the atomic gap toward the exact value. Hence, the m6-31G(d) basis set for Fc has been demonstrated to outperform many other popular and commonly used basis sets including TZV2P+f,[3] DZP,[8] DZVP,[33] LANL2DZ[9] and 6-31G*.[39]

The geometry of the optimized ferrocenium (Fc$^+$) and the effect of ionization on the geometry of the neutral molecule Fc (eclipsed) are compared with available theoretical and experimental studies in Table 1. In general, the impact of one electron ionization on the geometric properties of Fc$^+$ is very small with respect to Fc, in agreement with other studies.[2,28,39,40] For example, the bond lengths of the C-C and C-H bonds remain unchanged but with slightly longer distances for the iron related. The C-C and C-H bonds are 1.43 Å and 1.08 Å, respectively for Fc$^+$, which are actually the same as to the corresponding bond lengths in Fc calculated using the same model. However, the unrestricted B3LYP calculations show that the oxidation of Fc (i.e. Fc$^+$) expands the Fe related bond lengths. Similar results have been reported by earlier experiment[40] and previous theoretical studies.[2,28,39] Such the Fe-bond expansion implies that the electronic properties due to the removal of a bonded valence electron play the crucial role behind such changes in metal related bond lengths,[41-43] as in general cations would change in the opposite manner.

Table 1 also shows that the Fe-Cp and Fe-C distances of Fc$^+$ exhibit a similarly small increase of 0.02 Å with respect to the Fc counterparts, in agreement with most available studies. It is noted that two of the models, i.e., B3LYP/LANL2TZF and UHF/m6-31G(d) show reductions rather than expansions of the Fe relative bond distances, against other models in the same table. It is possibly due to insufficient electron correlation contained in the models, either the basis set (B3LYP/LANL2TZF) or the level of theory (UHF). Nevertheless, the calculations using different models are consistent except for the UHF/m-61G(d) model. For example, the Fe-C$_5$ and Fe-C bond distances of Fc$^+$ using the current UB3LYP/m6-31G(d) model are given by 1.69 Å and 2.09 Å, respectively, which is basically the same as the bond distance of 1.70 Å and 2.09 Å, calculated using the B3LYP/LANL2TZF model.[2]

As expected, that ionization of an electron from Fc to form $Fc^+$ leads to a small reduction of the molecular size (i.e. the electronic spatial extent $<R^2>$ as given in Table 1) of the ion. The size of $Fc^+$ decreases by 9.55 a.u., from 1358.84 a.u of $Fc^{10}$ to 1349.29 a.u. of $Fc^+$, which is the opposite with respect to the expansion of bond lengths upon ionization of Fc. It may be due to the following reasons. Firstly, the $Fc^+$ is one electron less than the neutral Fc, the coulomb force in $Fc^+$ is less strong than the same force in the neutral Fc. Hence, the bond lengths in $Fc^+$ can be longer. Secondly, it is also noted in Table 1 that the ∡$C_5$-H angle of the $Fc^+$ i.e. 1.23° is almost doubled than that was in neutral Fc system i.e. 0.66°,[10] such the angle increase will contribute to the expansion of the electron spatial extents. This is similar to two umbrellas connecting on their handles on Fe. In addition, the fact that only the Fe related bonds change in $Fc^+$ also indicates that the electron removed can be one of the electrons with strong correlation to the Fe atom originally.

*Ground electronic state configuration of ferrocenium ($Fc^+$)*

It is a known fact that electron correlation energy could alter ground electronic state configurations, so that the electron configuration of a molecule produced by the Hartree-Fock (HF) calculations are often not the same as those using the methods containing electron correlation energies such as MP2 and DFT methods. The same methods employed to study Fc and $Fc^+$ can enable one to make meaningful comparisons. In this section, the results obtained from the (U)B3LYP and (U)HF methods are discussed.

The ground electronic state of $Fc^+$ calculated using both the UBL3LP and UHF methods is $X^2A_1'$, rather than $X^2E_2'$ given by previous SCF calculations.[44,45] The present results are in agreement with an earlier study.[46] The total electronic energy of $Fc^+$ is -1650.40826860 $E_h$ using the UB3LYP/m6-31G(d) model, which inherits from corresponding neutral Fc conformer ($X^1A_1'$) with a total electronic energy of -1650.66192350 $E_h$ using the B3LYP/m6-31G(d) model.[14] When one β–electron is removed from the neutral Fc, the cation $Fc^+$ contains 48 α-electrons and 47 β-electrons. The core configuration of $Fc^+$ shows the same order in their core configuration of Fc,[14] so do the inner valence orbitals except for orbitals $(7a_1')^2$ and $(3e_1'')^4$ for both the α and the β electrons in $Fc^+$, which are swapped in $Fc^+$ with respect to their Fc. As a result, the UB3LYP calculation exhibit the following inner valence electron configuration for $Fc^+$ as

$(4a_1')^2 (3a_2'')^2 (3e_1')^4 (5a_1')^2 (4a_2'')^2 (2e_1'')^4 (4e_1')^4 (2e_2')^4 (2e_2'')^4 (6a_1')^2 (5a_2'')^2$

Fc $(X^1A_1')$: $\underline{(3e_1'')^4 (7a_1')^2} (5e_1')^4$

$Fc^+$ $(X^2A_1')$: $\mathbf{(7a_1')^2 (3e_1'')^4} (5e_1')^4$

The outer valence orbitals exhibit larger differences under this model as the related orbitals are not simply swapped. In $Fc^+$, although the α-electrons and β-electrons are calculated separately in the unrestricted calculations, the electron configurations are the same, except the position of the singly occupied α-orbital $(\mathbf{8a_1'})$ which is missing (becomes the LUMO) in β–electron of $Fc^+$. As a result, the complete electronic configurations of $Fc^+$ is given by the UB3LYP/m6-31G(d) model as:

Fc ($X^1A_1'$): (core) ... $(3e_2')^4 (3e_2'')^4 (6a_2'')^2 (4e_1'')^4 (6e_1')^4 \underline{\mathbf{(8a_1')^2}} \mathbf{(4e_2')^4} (5e_1'')^0$

Fc$^+$ ( X$^2$A$_1$'): (core) ... (3e$_2$')$^4$ (3e$_2$")$^4$ (6a$_2$")$^2$ **(8a$_1$')**$^1$ (4e$_1$")$^4$ (6e$_1$')$^4$ **(4e$_2$')$^4$** (5e$_1$")$^0$

It is important to note that the singly occupied molecular orbital (SOMO) of Fc$^+$ is NOT the highest occupied molecular orbital (HOMO) in Fc nor in Fc$^+$. Orbital 8a$_1$' is the HOMO-1 in Fc, which becomes the SOMO in Fc$^+$ but it is in fact the fourth highest occupied molecular orbitals (i.e., HOMO-3) in the α electron configuration, whereas this orbital (8a$_1$') is absent in β electron irrespective to the methods. Such the significant change in orbital configuration indicates that the ionization of a β–electron from Fc is not a small perturbation, but a profound change of the electronic structure of ferrocene.

*Impact on the valence configuration of Fc$^+$ upon removal of a β-electron from Fc*

To better understand the electron configuration of Fc$^+$, parallel calculations based on the (U)HF/m6-31G(d) model have also been performed. Both the (U)B3LYP/m6-31G(d) and (U)HF/m6-31G(d) results show the valence electron configurational differences the of α and β electrons, respectively, between Fc and Fc$^+$. Figure 1 provides the calculated outer valence orbital energy diagrams of the Fc$^+$ with respect to the Fc counterpart, using the UB3LYP/m6-31G(d) (Fig 1(a)) and UHF/m6-31G(d) models (Fig 1(b)). For neutral Fc, the configurations (i.e., the order of the outer valence orbitals) vary with the methods employed[34] and the calculations show that the Fe d-electrons dominate the frontier orbitals (e.g., Fe-d$_z^2$ for 8a$_1$'; Fe-d$_{x^2-y^2}$, and Fe-d$_{xy}$ for the doubly degenerate orbitals 4e$_2$').

Ionization of a (β-) electron from Fc changes the electron configurations of Fc$^+$. Interestingly, the doubly degenerate and doubly occupied HOMO (4e$_2$') of Fc remains as the same doubly degenerate and doubly occupied HOMO ($4e_2'$) in $Fc^+$ (Fig 1(a)). When a β-electron is removed from the neutral Fc, the α-electrons of the $Fc^+$ retain the same configuration as the neutral Fc except that the local orbital changes significantly, both in the α– and β–configurations. That is, the orbital energy position of the SOMO $8a_1'$ (α) shifts down from HOMO-1 (MO46) in Fc to HOMO-3 (MO42) in $Fc^+$, that is, becomes the first (highest) singly occupied orbitals the three pairs of doubly degenerate MOs of $Fc^+$. The same as the β-electrons of $Fc^+$ but the orbital $8a_1'$ becomes the lowest unoccupied molecular orbital (LUMO) of the $Fc^+$ in the β-electron configuration. The LUMO - HOMO energy gap of Fc ($5e_1''$ - $4e_2'$) calculated using the B3LYP/m6-31G(d) is given by 5.30 eV (see Figure 1 (a)), is close to the gap of 5.42 eV ($D_{5h}$) using the PW91/TZ2P model of Zhang et al.[64] In the α-electron diagram (middle column of Figure 1(a)) of $Fc^+$, this HOMO - LUMO energy gap is reduced to 5.03 eV using the unrestricted calculations. Such the large HOMO-LUMO energy gaps in Fc and α-$Fc^+$, as a result, contribute to the low spin configuration of Fc, in agreement with other studies.[25,34] It is noted that the α-electron in SOMO ($8a_1'$) is neither the HOMO nor the HOMO-1 of $Fc^+$, instead, this SOMO becomes the 4$^{th}$ MO, i.e., HOMO-3 in the α-electron configuration of $Fc^+$, whereas the order other outer valence electrons remains the same as if they were in the neutral Fc. Perhaps the largest difference between the three electron configurations (Fc, $Fc^+$-α and $Fc^+$-β) in this Figure 1(a) is that the LUMO ($8a_1'$) of the β-electrons of $Fc^+$, i.e., the orbital from where an electron is removed, is different from the LUMOs of Fc ($5e_1''$) and α–$Fc^+$ ($5e_1''$). Such the ground electron configurations of Fc and $Fc^+$ given in Figure 1(a) will certainly lead to a complex first ionization process for Fc.

Electron correlation effects of Fc and $Fc^+$ are significant. Figure 1(b) gives the orbital diagrams of Fc and $Fc^+$ using the (U)HF/m6-31G(d) model. As shown in Figure 1(a) and (b), the most significant differences are the outer valence electron configurations and the HOMO-LUMO energy gap. The configurations of the frontier orbitals of the same species between the (U)B3LYP/m6-31G(d) and (U)HF/m6-31G(d) are very different, as discovered in neutral Fc,[14] which is also the case in the cation $Fc^+$. Interestingly, the $\alpha$-$Fc^+$ exhibits significant configurational changes but the $\beta$–$Fc^+$ experiences less significant configurational changes in the outer valence space from their (U)HF/m6-31G(d) calculations. For example, the $8a_1'$ orbital of Fc locates at the HOMO-4 position but moves to HOMO-5 position in the $\alpha$-$Fc^+$. The next significant electron correlation effect is the HOMO-LUMO energy gaps. Such the energy gaps were calculated as 5.30 eV, 5.03 eV and 2.90 eV for Fc, $\alpha$-$Fc^+$ and $\beta$–$Fc^+$, respectively, using the (U)B3LYP/m6-31G(d); whereas the HOMO-LUMO energy gaps of the same species calculated using the (U)HF/m6-31G(d) are 13.42 eV, 9.06 eV and 13.78 eV, accordingly. It indicates that the studies of electronic structures of Fc and $Fc^+$ without proper inclusion of electron correlation effect may lead unreliable results.

Although the outer valence electron configurations of the neutral Fc are quite different using the models with and without electron correlation, what is in common is that both models, UB3LYP/m6-31G(d) and UHF/m6-31G(d), indicate that in $Fc^+$ the $\beta$-electron is removed from the $8a_1'$ orbital of the neutral Fc, and this orbital in the $\beta$-electron diagram (right column of Figure 1(a & b)) becomes the lowest unoccupied molecular orbital (LUMO), in agreement with the UB3LYP/m6-31G(d) calculations.

*Impact on α- and β-electrons upon ionization of Fc: EOES*

Electron configuration reflects the order (configuration) of the orbital energies and orbital symmetry of a compound. It is agreed from earlier studies[47-49] that the ionization of ferrocene causes a large relaxation in the electronic structure of ferrocenium with significant relaxation energy. The next question is which orbital or orbitals are the responsible ones in the ionization of Fc? To this extend, the excess orbital energy spectrum (EOES)[14] is able to provide detailed and orbital based quantitative measurement to monitor such the changes. As a result, the EOES between Fc and Fc$^+$ is employed in the present study to reveal more insight.

Figure 2(a) compares the EOES of the α–electrons (in black) and β–electrons (in red) of Fc$^+$ with respect to the counterparts in neutral Fc using the (U)B3LYP/m6-31G(d) model. That is, the EOES reveal the (orbital symmetry correlated) exceed orbital energy differences between the cation Fc$^+$ and neutral Fc, i.e.,

$$\Delta\varepsilon_i^\alpha = \varepsilon_i^\alpha (Fc^+) - \varepsilon_i (Fc)$$

and

$$\Delta\varepsilon_i^\beta = \varepsilon_i^\beta (Fc^+) - \varepsilon_i (Fc)$$

The EOES in Figure 2(a) indicates that the α–electrons and β–electrons of Fc$^+$ are significantly different from Fc. Interestingly, Figure 2(a) shows that larger impact on the α–electrons (black spectrum) rather than the β–electrons (red spectrum) upon ionization of Fc/Fc$^+$. In general, the impact of ionizing one electron of Fc causes energy changes in ALL orbitals, α-electrons and β-electrons, core and valence of the complex. All such orbital energy changes are larger (more negative) than -110 kcal·mol$^{-1}$, regardless of α-electrons and β-electrons. It is the α-electrons of Fc$^+$ which exhibit more significant orbital energy changes from their Fc counterparts, rather than the β-electrons with one less electron. For example, all significant valence energy changes with more (negative) than -155 kcal·mol$^{-1}$ are from the α–electrons, that is, the black spectrum is below the red spectrum in Figure 2(a). Similar comparisons of the EOES using both the (U)B3LYP/m6-31G(d) and the (U)HF/m6-31G(d) models are given in Figure S1 of the supplementary materials.

The EOES in Figure 2 also indicates that the Fe-electrons exhibit significantly stronger correlation than the electron correlation of other atoms in Fc i.e. C and H atoms. Similar to the previous EOES of Fc eclipsed and staggered conformers,[14] the large excess orbital energies concentrate in three clusters of Fe-related orbitals, that is, the core region of MO1-5, inner valence region of MO16-19 as well as outer valence region in particular MO46-48 (where MO47-48 are the doubly degenerate HOMOs). A further inspection indicates that those Fe-dominant orbitals show a common pattern (v - -) in a series of three orbitals with a larger negative energies followed by a less negative and degenerate pair of α orbitals. The three boxes marked as A, B and C in Figure 2(a) indicate such the excess energy pattern of (v - -). Three core α-orbitals in box A (MOs 3-5) are again, dominated by Fe-2$p_z$ and (Fe-2$p_x$, Fe-2$p_y$) electrons; the next box B (MOs 17-19) in the inner valence region are dominated by Fe-3$p_z$ and (Fe-3$p_x$, Fe-3$p_y$) electrons. More interestingly, the last box C in the outer valence region is dominated by, in fact, the Fe-d-electrons, Fe-3$d_z^2$ and (Fe-3$d_{x^2-y^2}$, Fe-3$d_{xy}$).

A common feature of the three MOs in boxes A, B and C, respectively, is the Fe z-electron domination, followed by the doubly degenerate electrons of the Fe x- and y-electrons. However, the inner valence α electrons i.e. Box B are more different in energy than either the respective core or outer valence electrons, i.e. Box A and C. Among the z-electrons in these zones, the excess energy order is given by (negative) $\Delta\varepsilon$(Fe-2p$_z$, ~ 170 kcal·mol$^{-1}$) < $\Delta\varepsilon$(Fe-3d$_z^2$ ~195 kcal·mol$^{-1}$) < $\Delta\varepsilon$ (Fe-3p$_z$ ~ 185 kcal·mol$^{-1}$), as seen in this figure. Such the excess energies of the Fe x- and y-electrons are similar, $\Delta\varepsilon$(Fe-2p$_x$, Fe-2p$_y$ ~150 kcal·mol$^{-1}$) < $\Delta\varepsilon$(Fe-3d$_z^2$ ~158 kcal·mol$^{-1}$) < $\Delta\varepsilon$ (Fe-3p$_x$, Fe-3p$_y$ ~150 kcal·mol$^{-1}$), for the doubly degenerate core, outer valence and inner valence orbital pairs, accordingly. The large excess orbital energies in the three clusters A, B and C of the α-electrons indicate that it is not merely an outer valence event to remove a β-electron in the outer valence space of Fc, and the Fe electrons correlate more significantly than electrons from Cp ligand in Fc and its cation.

In the open shell cation, Fc$^+$, there are also very large excess orbital energies between the α- and β-electrons for the relaxation energy. Figure 2(b) illustrates EOES for the α- and β-electrons of Fc$^+$. Again, a couple of interesting features can be seen from this spectrum. First, the impact of the ionization process is larger than relaxation of α- and β-electrons of Fc$^+$ although the latter is considerably large for Fe-dominant orbitals. As shown in the EOES in Figure 2(b), such the relaxation energy for other orbitals is very small or negligible except for the Fe-dominant orbitals. The largest orbital energy relaxation among the α and β electrons of Fc$^+$ exceed 60 kcal·mol$^{-1}$ in the inner valence shell. Although a β-electron is removed from Fc for the ionization, it is the β–electrons rather than the α-electrons of Fc$^+$ whose orbital energies are less relaxed from their neutral Fc counterparts.

*Ionization of Fc and its fingerprint orbital 8a$_1$' of ferrocenium*

Removal of one bonding electron from Fc leads to its cation, i.e., ferrocenium (Fc$^+$). The debate from which orbital of Fc the electron is removed to form Fc$^+$, has been as much as the debate which conformer of ferrocene is more stable. As shown by the results obtained in the previous sections, the strong electron correlation among the Fe-electrons of Fc and $Fc^+$, core and valence, suggests that it is unlikely to correctly work out the electronic structures of Fc and $Fc^+$ if the model did not include sufficient electron correlation energy and relaxation energy. For example, a number of earlier studies using HF (SCF) model reported[45,50-53] claimed that the electron is removed from degenerated $4e_2'$ ($4e_{2g}$) orbitals, thereby suggesting that the ground state configuration of $Fc^+$ is $X^2E_2'$ $[(e_2')^3 (a_1')^2]$[45] or $X\ ^2E_{2g}$ $[(e_{2g})^3 (a_{1g})^2]$[50-53].

In Figure 2, the significant orbital electron energy changes at the ionization reveal that the orbital relaxation effect can be too large to ignore. That is, the Koopman's frozen orbital approximation break down in this case. Rather, the ionization energy must be calculated using the total energy difference between Fc and $Fc^+$ using the $\Delta$ method as,

$$IP = E\ (Fc^+) - E\ (Fc)$$

Table 2 compares the calculated IP of Fc with prior high level theoretical[2] and experimental results.[46,54-63] The present calculated IP of Fc ($D_{5h}$) is in excellent agreement with prior studies[2,46,54-63]. The present calculated IP of eclipsed Fc is 6.90 eV whereas the experimentally measured values in the literature vary in the range of 6.71-6.99 eV below 7 eV, depending on the techniques used.[54-63] For example, the present $\Delta$DFT method yields the IP as 6.90 eV (and 6.89 eV for staggered conformer) using the (U)B3LYP/m6-31G(d) calculations, which agrees well with the most recent electron impact measurement of 6.99 eV.[55] A number of other measurements such as electron impact[46,59,60] and photoelectron spectrum studies[62] produced the IP value of 6.90 eV, which is almost identical to the present calculated IP of Fc. In addition, it is not very surprise that the $\Delta$SCF (HF) results of 6.85 eV (6.86 eV for $D_{5d}$) are also very good agreement with measurements---almost as good as high level methods. It can be the indication that relaxation energy is critically important whereas the electron correlation energy between Fc and $Fc^+$ are very similar so that they cancelled out.

The equation brings the ionization potential (IP) of Fc close using the (U)HF and/or (U)B3LYP models, as the electron correlation energies are cancelled out in the $\Delta$ method. The (U)B3LYP/m6-31G(d) model reveals that the ground electronic configuration ($X^2A_1'$) of $Fc^+$ is $(8a_1')^1 (4e_2')^4 (5e_1'')^0$, which is in well agreement with the ground electronic state of $Fc^+$ reported by the Begun and Compton[46]. As pointed by Haaland[20], if one defines the HOMO of Fc as those from which electrons are most easily removed, the experimental evidence points to $4e_2'$ and $8a_1'$ orbitals, in agreement with the orbital energies calculated in the present study.

The $8a_1'$ orbital is unique in Fc and $Fc^+$ and therefore, it is the signature orbital of ionization of Fc. Figure 1 and 2(a) show that a small number of non-degenerate orbitals experience particularly large excess orbital energies in $Fc^+$. In the valence space, three $\alpha$-orbitals, which exhibit larger excess orbital energies than the IP (i.e. 160 kcal·mol$^{-1}$) upon ionization, that is, $4a_1'$ (MO16), $3a_2''$ (MO17) and $8a_1'$ (MO46 for Fc or MO42 for $\alpha$-$Fc^+$), are non-degenerate orbitals with *a*-irreducible representative. To further understand this cluster of signature orbitals ($4a_1'$, $3a_2''$ and $8a_1'$) which experience large impact in in the ionization process, Figure 3 compares the ($\alpha$-)electron charge distributions of those orbitals of $Fc^+$ (upper row) with respect to their neutral Fc (lower row) counterparts. These signature orbitals, $4a_1'$, $3a_2''$ and $8a_1'$ (for $\alpha$-$Fc^+$), are all Fe-related orbitals dominated by Fe-3s, Fe-3$p_z$ and Fe-3$d_z^2$ orbitals, respectively. The Fe-d electron dominance in the orbital $8a_1'$ agrees with Scuppa et al.[34] A common property of these orbitals is that they are aligned along the directions of the z-axes, i.e. the Cp-Fe-Cp axis. Although the energies of these orbitals change significantly as shown by the EOES in Figure 2, the inner valence orbitals of $4a_1'$ (MO16) and $3a_2''$ (MO17) exhibit very similar electron density distributions between $Fc^+$ and Fc as seen in Figure 3. However, the electron distributions of the $8a_1'$ orbitals of $Fc^+$ ($\alpha$) and Fc shows some subtle differences in the contributions of the pair of the Cp rings.

Figure 4 presents the orbital theoretical momentum profiles (TMP) for the three signature orbitals. The TMDs of orbitals $4a_1'$, $3a_2''$ and $8a_1'$ provide additional information of the orbitals in momentum space quantitatively from the information obtained from the coordinate space through a Fourier transform, an analysis called using dual space analysis (DSA).[26] The electrons in $4a_1'$ and $3a_2''$ have almost identical orbital momentum profiles in $Fc^+$ ($\alpha$, in red) and in Fc (in black). The half bell shaped orbital profiles of $4a_1'$ (MO16) confirms that it is an s-electron (without any nodal plane) dominant orbitals (i.e., Fe 3s), whereas the bell shaped orbital profiles for $3a_2''$ (MO17) indicate that the orbital is associated with an orbital of a nodal plane, i.e., p-electron dominant orbitals (i.e., Fe $3p_z$). The last orbital profiles of $8a_1'$ are very interesting, which exhibit significantly quantitative different d-electron dominant component (Fe $3d_z^2$). The max-min-max shape profiles indicate a d-electron dominant orbital profile of $8a_1'$. Significant differences in the small momentum region of $p < 1.5$ a.u. of $8a_1'$ orbital profile in Figure 4, indicate that the long range region of the orbitals ($8a_1'$) in Fc and $Fc^+$ are very different. The orbital momentum profile is a very sensitive property to identify such the difference. Therefore, we define this singly occupied $\alpha$-electron orbital of $8a_1'$ as the fingerprint orbital of ferrocenium.

## 4. Conclusions

The present study employs the excess orbital energy spectrum (EOES) of eclipsed $Fc^+$ to study the changes of the α-electrons and in the β-electrons of ferrocenium ($Fc^+$) from neutral ferrocene (Fc), as well as the ionization of Fc with respect to the singly occupied α-orbital of $Fc^+$. Density functional theory (DFT) based (U)B3LYP/m6-31G(d) and the ab initio (U)HF/m6-31G(d) models are used in the calculations. Similar to the previous EOES study on the E-Fc and S-Fc conformers of Fc,[14] the present EOES of Fc and $Fc^+$ indicates that electrons of the transition metal Fe from both core and valence region exhibit significant orbital energy changes due to the impact of removal of one electron from Fc upon ionization, suggesting that the Fe-electrons are more strongly correlated than electrons from the ligand such as Cp. The present study discovers that it is the α-electrons (not the β-electrons from where an electron is removed) which experience more significant changes with respect to Fc upon ionization, and identifies three Fe-dominant orbital clusters, in core (MO1-5), inner valence (MO16-19) and outer valence spaces (MO42-48), show certain pattern of changes with respect to ionization. The electrons of the metal, Fe, are more strongly correlated in Fc and $Fc^+$ with significant correlation energy. The present study further reveals that significant orbital relaxation upon ionization makes the frozen orbital Koopman's theorem inaccurate. As a result, the ionization potential can only be accurately calculated using the ΔDFT (6.90 eV) or ΔSCF (6.85 eV) methods to achieve excellent agreement with measurements (6.90 eV[46,59,60,62] or 6.99 eV[55]) and electron correlation is largely cancelled out. The calculations suggest that the β-electron of Fc unlikely to be not removed from the HOMO ($4e_2'$) of Fc, rather, this electron is removed from the HOMO-1 ($8a_1'$) orbital of Fc, supporting the statement of Haaland[20], that the "HOMO" of Fc is the orbital from which electrons are most easily removed. Finally, dual space analysis (DSA)[26] using orbital density and orbital momentum profiles determines that the SOMO α-($8a_1'$) orbital of $Fc^+$ as the fingerprint orbital of $Fc^+$ from the signature orbitals of $4a_1'$, $3a_2''$ and $8a_1'$(α only), as this orbital ($8a_1'$) experiences significant changes not only in their energies but also in its momentum profiles.

**Acknowledgements**

Swinburne University Supercomputing Facilities must also be acknowledged. SI acknowledges Swinburne University Postgraduate Research Award (SUPRA).

**Table 1: Geometric properties of Fc$^+$ using different models** ($\Delta$= Fc$^+$ - Fc).[a]

| Parameter | UB3LYP/ m6-31G*[b] | B3LYP/ 6-31+G*[28] | B3LYP/ LANL2TZF[c] | B97-D/ 6-31+G*[28] | UHF/ m6-31G*[b] | MP2/ 6-31+G*[28] | Expt[40] |
|---|---|---|---|---|---|---|---|
| Fe-C$_5$ (Å) | 1.69 (0.02) | 1.70 (0.02) | 1.70 (-0.02) | 1.71 (0.08) | 1.80 (-0.05) | 1.54 | 1.68 (0.02) |
| Fe-C (Å) | 2.09 (0.02) | 2.09 (0.02) | 2.09 (-0.02) | 2.10 (0.06) | 2.17 (-0.04) | 1.97 | 2.069 |
| C-C (Å) | 1.43 (0.0) | 1.43 (0.0) | 1.43 (0.0) | 1.44 (0.0) | 1.41 | 1.44 | |
| C-H (Å) | 1.08 (0.0) | 1.08 (0.0) | 1.08 (0.0) | 1.09 (0.0) | 1.07 | 1.08 | |
| ∡C$_5$-H (°) | 1.23 | 1.60 | | 1.80 | 0.29 | 1.60 | |
| <R$^2$>(a.u.) | 1349.29 | | | | 1438.74 | | |

[a] The variations with respect to the corresponding Fc conformer using the same model are given in parenthesis.
[b] This work (m6-31G* for Fe, 6-31G* for C and H)
[c] Basic Set : LANL2TZF for Fe, 6-31G* for C and H (Ref [2])

**Table 2: Comparison of ionization potential (IP) of Fc with literature values.**

| Methods | IP (eV) |
|---|---|
| **Theory** | |
| B3LYP/m6-31G(d)[a] | 6.90 (6.89) |
| HF/m6-31G(d)[a] | 6.85 (6.86) |
| G3(MP2)-RAD | 7.062[2] |
| G3(MP2)-RAD-Full/6-31G(d)/LanL2TZf[b] | 7.067[2] |
| G3(MP2)-RAD-Full/TZ[b,c] | 7.047[2] |
| **Expt** | |
| PIMS[d] | 6.747± 0.009[54] |
| EIMS[e] | 6.99[55], 6.9±0.2[59], 6.9±0.1[46,60] |
| N/A | 6.71±0.08[57] |
| CTE (Pulsed MS)[f] | 6.81± 0.07[56] |
| ETEC[g] | 6.82[58] |
| PES[h] | 6.72[61], 6.90[62] |
| CTS[i] | 6.97[63] |

[a] This work (m6-31G* for Fe, 6-31G* for C and H) $D_{5h}$ and IP for $D_{5d}$ in parrnteses.
[b] Additional core correlation using the (RO)CCSD(T,Full)/6-31G(d)/LanL2TZf level of theory
[c] includes core correlation corrections using (RO)CCSD(T)/6-311+G(d,p) as theory
[d] Photoionization Mass Spectroscopy
[e] Electron impact Mass Spectroscopy
[f] Charge transfer equilibrium
[g] Electron transfer equilibrium constant (FT-ion resonance MS)
[h] Photoelectron spectroscopy
[i] Charge transfer spectra

**Figure 1: Outer valence orbital energy diagrams of Ferrocene and Ferrocenium using (a) (U)B3LYP/m6-31G(d) and (b) (U)HF/m6-31G(d) models.**

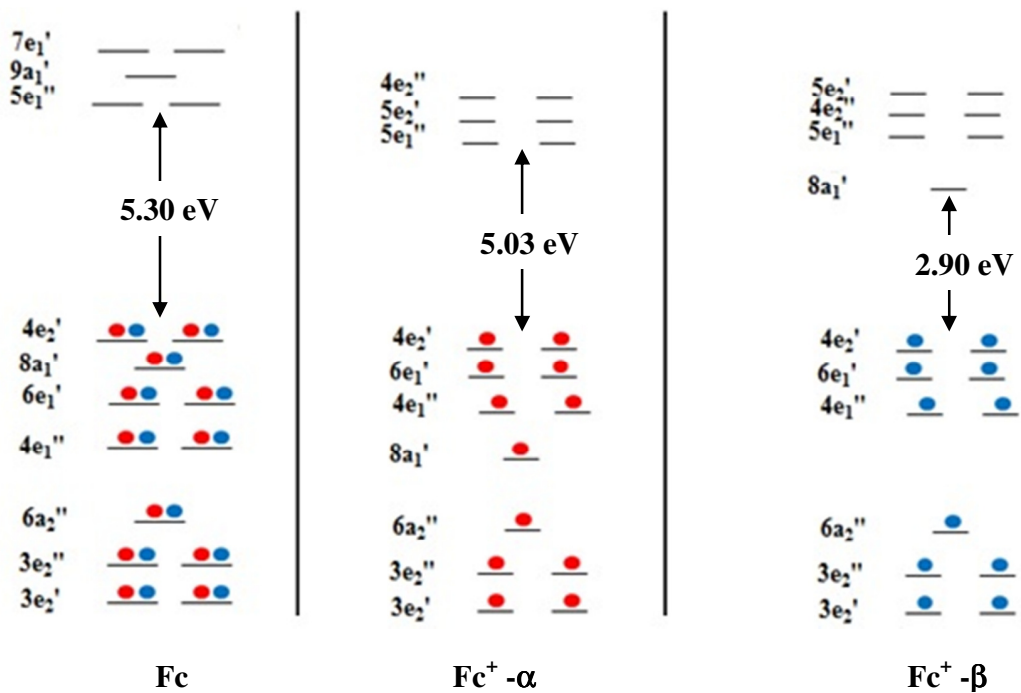

(a) UB3LYP/m6-31G(d)

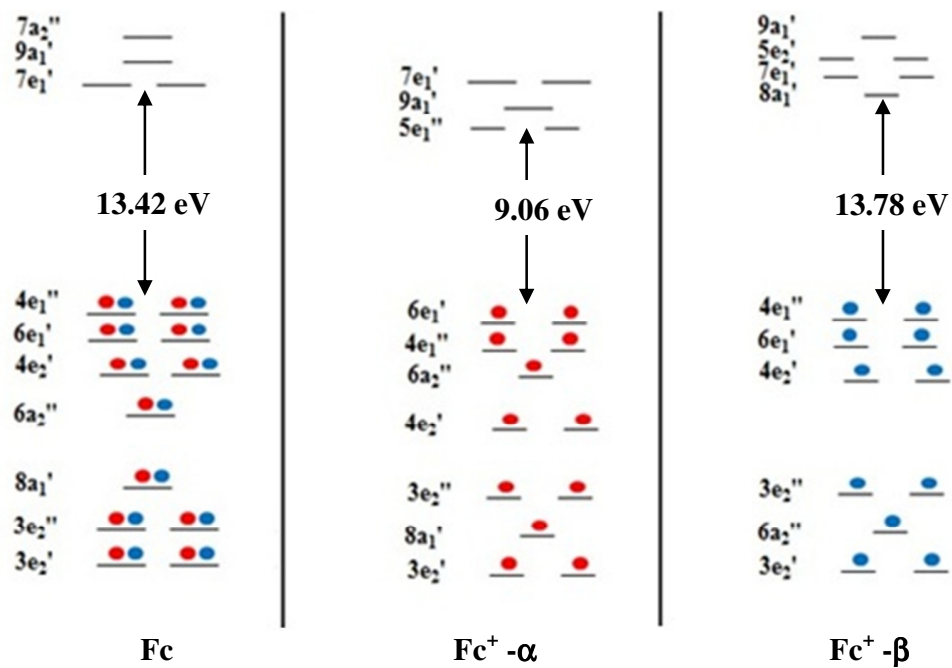

(b) UHF/m6-31G(d)

## Figure 2

**(a)** EOES of Fc$^+$ ($\alpha$–electrons ---black and $\beta$–electrons – red) with respect to Fc.

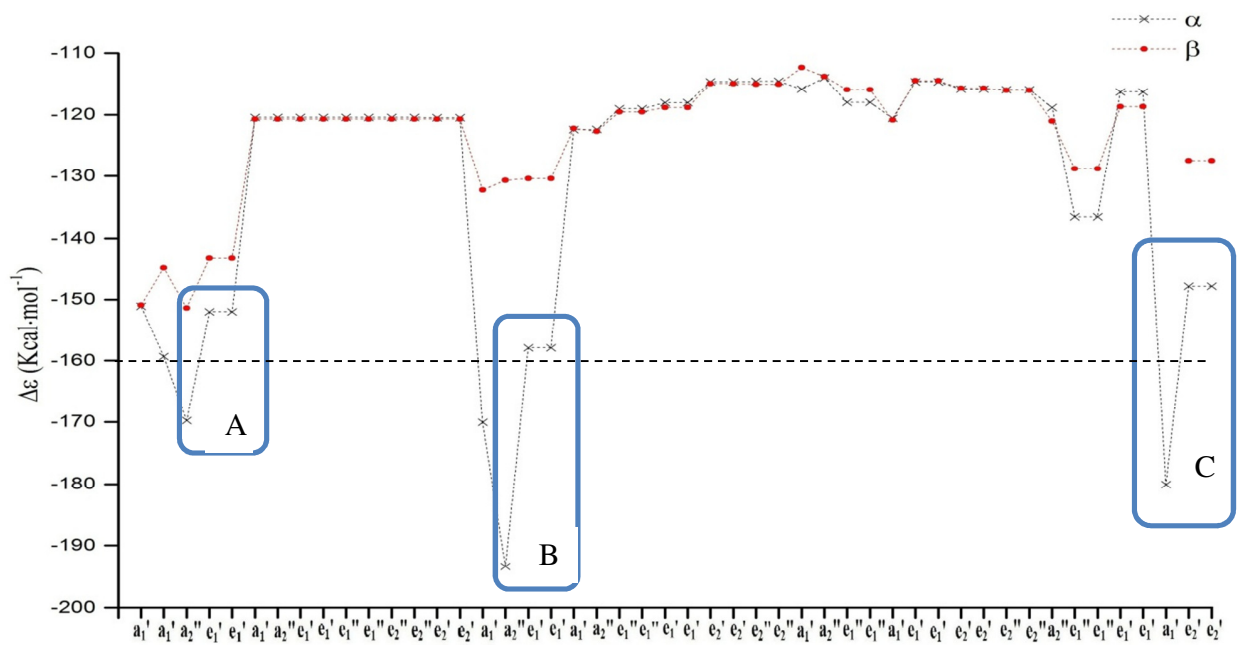

(a) EOES of $\alpha$–electrons and $\beta$-electrons of Fc$^+$.

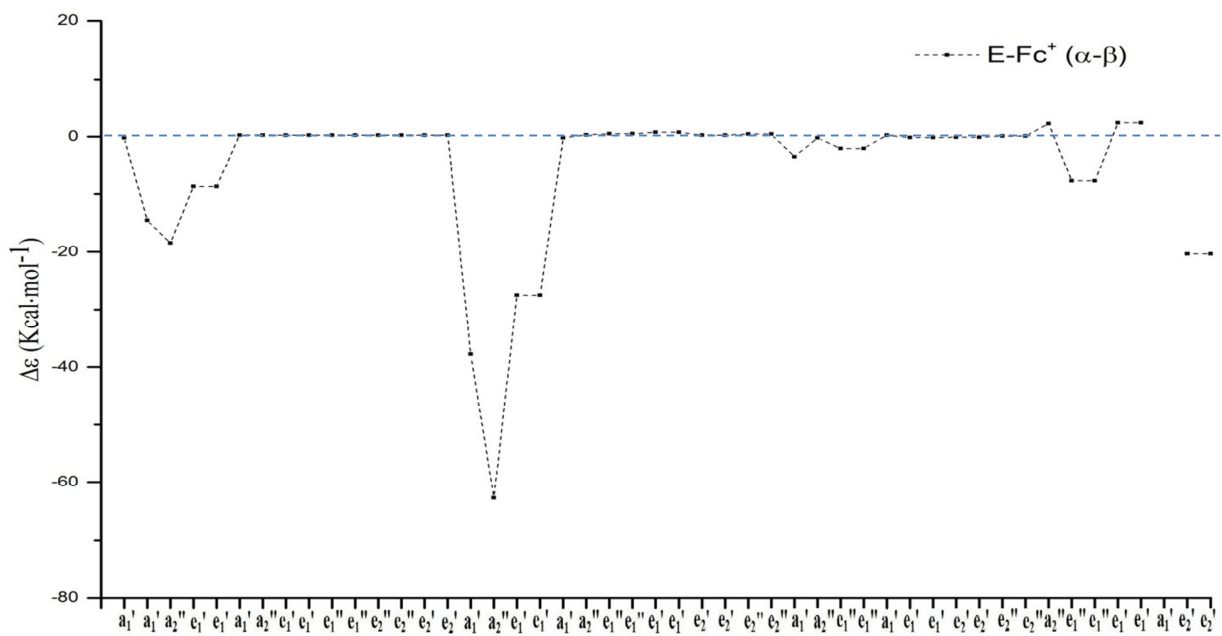

**Figure 3:** Comparison of the signature orbitals ($4a_1'$, $3a_2''$ and $8a_1'$) experience large changes among ($\alpha$-) electrons of Fc+ (Upper panel) and Fc (Lower Panel).

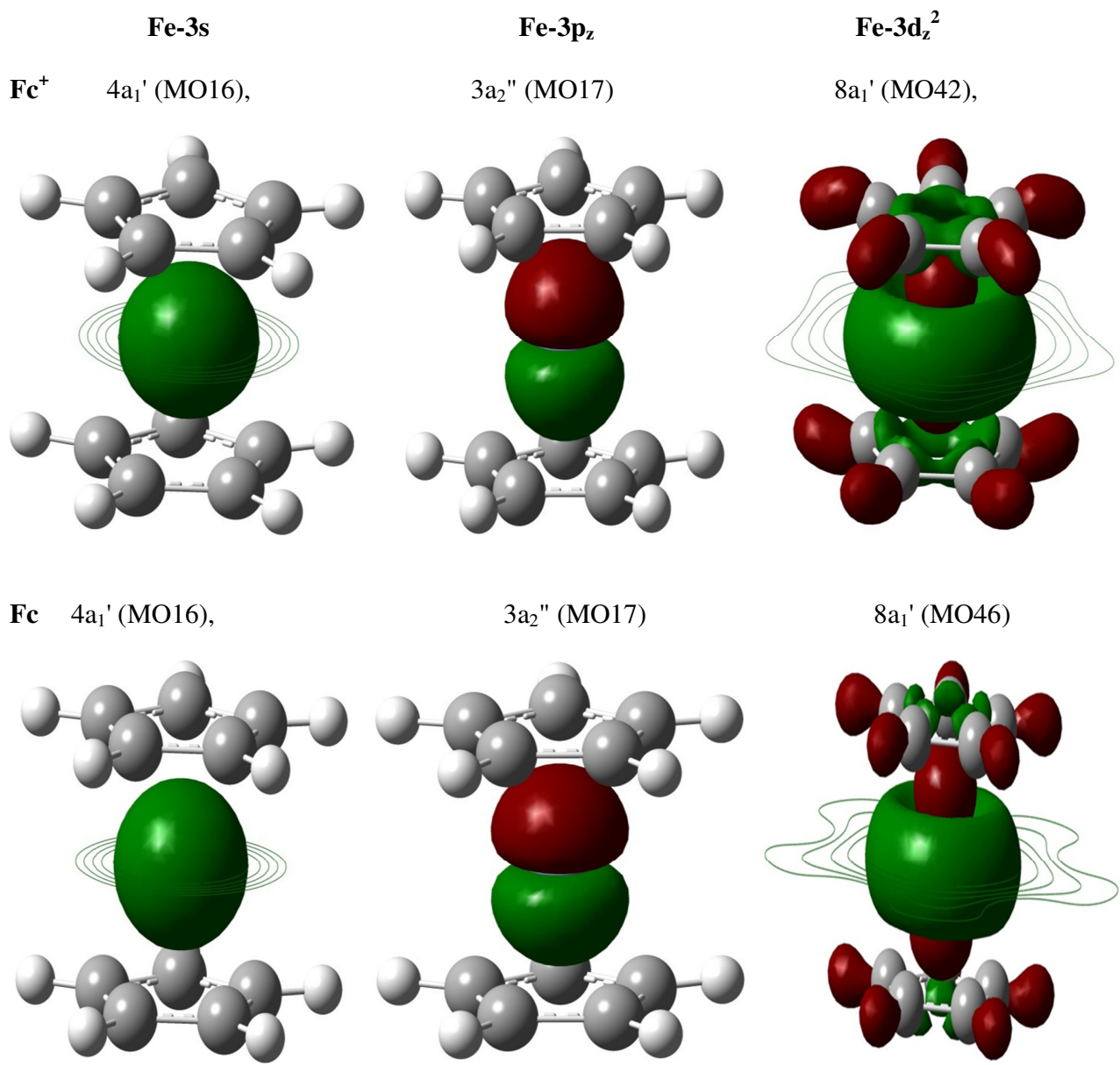

|  | **Fe-3s** | **Fe-3p$_z$** | **Fe-3d$_z^2$** |
|---|---|---|---|
| **Fc$^+$** | $4a_1'$ (MO16), | $3a_2''$ (MO17) | $8a_1'$ (MO42), |
| **Fc** | $4a_1'$ (MO16), | $3a_2''$ (MO17) | $8a_1'$ (MO46) |

**Figure 4:** Theoretical momentum distribution (TMD) profile of the signature orbitals $4a_1'$, $3a_2''$ and $8a_1'$ of Fc and $Fc^+$ in which the last orbital, $8a1'$ is the fingerprint orbital of Fc+ due to the change of orbital momentum profile. Here α-electrons of $Fc^+$ (red) and Fc (black) (Note that the fingerprint orbital $8a_1'$ is SOMO).

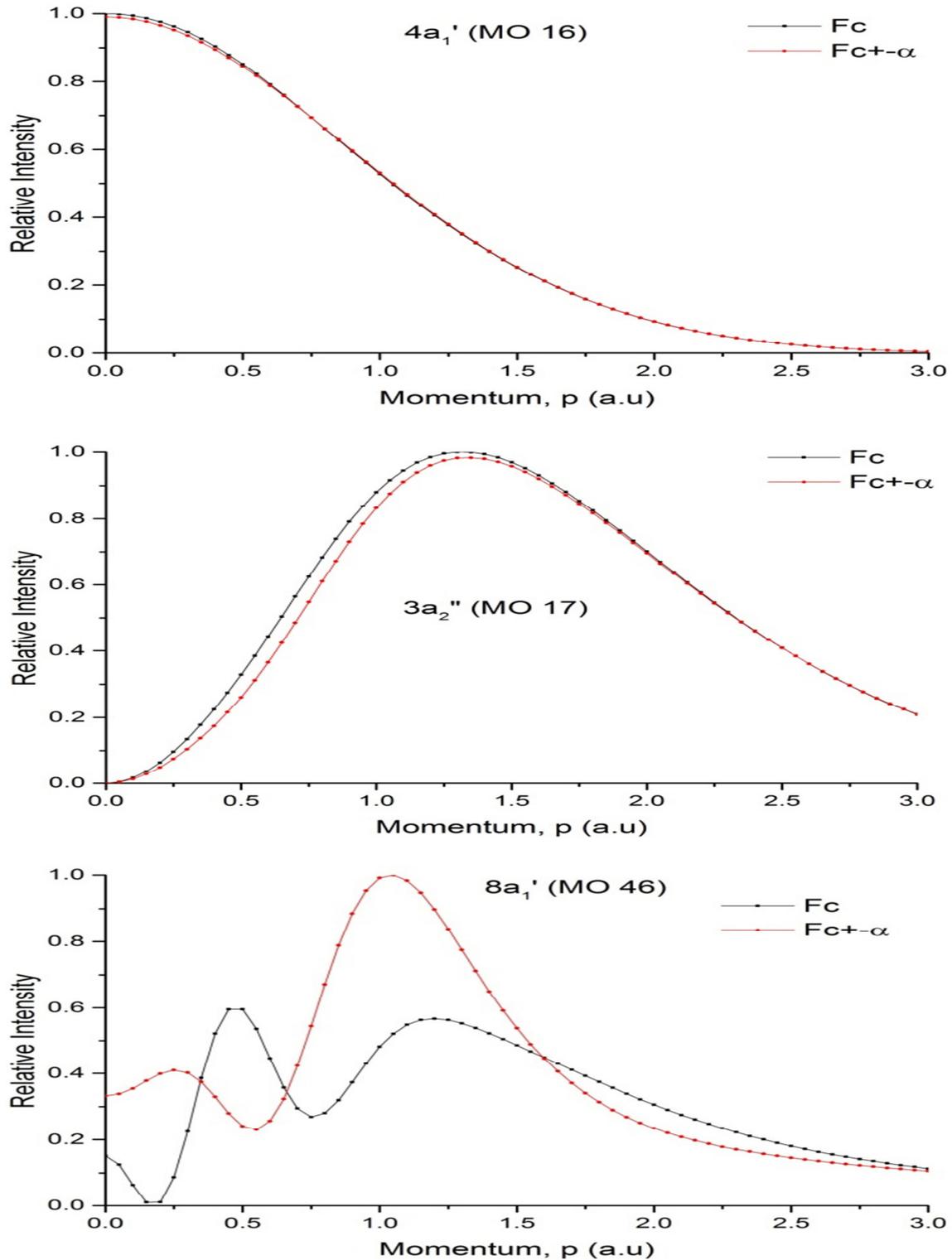

**Figure S1:** The EOES of eclipsed ferrocenium (α,β) with respect to ferrocene using (U)B3LYP/m6-31G(d) and (U)HF/m6-31G(d) models.

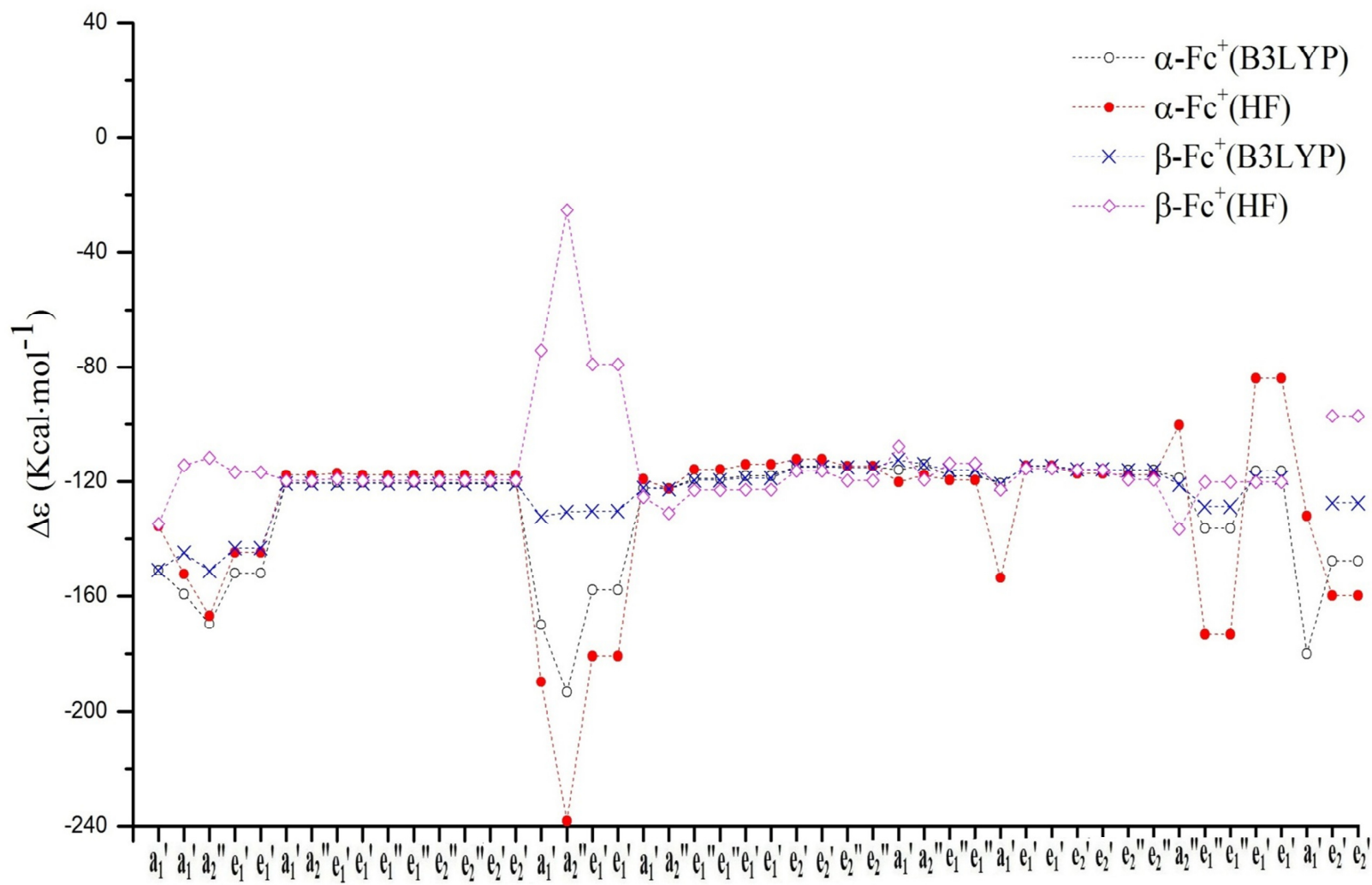